\documentclass{an}
\usepackage{graphicx}
\usepackage{times}
\Volume{01}              
\Year{2002}              
\Month{10}               
\Pagespan{223}{227}      

\begin{document}
\lhead[\thepage]{D.~M. Alexander et~al.: The 2~Ms Chandra Deep Field-North Survey}
\rhead[Astron. Nachr./AN~{\bf XXX} (200X) X]{\thepage}
\headnote{Astron. Nachr./AN {\bf 32X} (200X) X, XXX--XXX}

\title{Resolving the Source Populations that Contribute to the X-ray Background: The 2~Ms Chandra Deep Field-North Survey}

%
%




\author{D.~M. Alexander, F.~E.~Bauer, W.~N.~Brandt,
G.~P.~Garmire, \hbox{A.~E.~Hornschemeier}, D.~P.~Schneider, and
C.~Vignali}

\institute{Department of Astronomy \& Astrophysics, 525 Davey Lab, 
The Pennsylvania State University, University Park, PA 16802.}

\date{Received {\it date will be inserted by the editor}; 
accepted {\it date will be inserted by the editor}} 

\abstract{
With $\approx$~2 Ms of exposure, the Chandra Deep Field-North (CDF-N)
survey provides the deepest view of the Universe in the 0.5--8.0 keV
band. Five hundred and three (503) X-ray sources are detected down to
on-axis 0.5--2.0~keV and 2--8~keV flux limits of $\approx 1.5\times
10^{-17}$~erg~cm$^{-2}$~s$^{-1}$ and $\approx 1.0\times
10^{-16}$~erg~cm$^{-2}$~s$^{-1}$, respectively. These flux limits
correspond to $L_{\rm 0.5-8.0~keV}\approx3\times10^{41}$ erg~s$^{-1}$
at $z=$~1 and $L_{\rm 0.5-8.0~keV}\approx2\times10^{43}$ erg~s$^{-1}$
at $z=$~6; thus this survey is sensitive enough to detect starburst
galaxies out to moderate redshift and Seyfert galaxies out to high
redshift. We present the X-ray observations, describe the broad
diversity of X-ray selected sources, and review the prospects for
deeper exposures.
}

\correspondence{davo@astro.psu.edu}

\maketitle

%
\section{Introduction}
%

One of the driving goals behind the construction of the {\it Chandra
X-ray Observatory\/} (hereafter {\it Chandra}; Weisskopf et~al. 2000)
was to perform the deepest possible X-ray studies of the Universe.
Great advances in this direction have been made with the completion of
two $\approx$~1~Ms surveys [the Chandra Deep Field-North (CDF-N);
Brandt et~al. 2001a, and the Chandra Deep Field-South (CDF-S);
Giacconi et~al.  2002]. These ultra-deep {\it Chandra} surveys resolve
the bulk of the 0.5--8.0~keV background, providing the deepest views
of the Universe in this band (e.g.,\ Campana et~al. 2001; Cowie
et~al. 2002) and uncovering a broad variety of source populations (see
Brandt et~al. 2002 for a review).

The CDF-N was awarded a second $\approx$~1~Ms of {\it Chandra}
exposure in Cycle 3. Here we provide a brief overview of the 2~Ms
\hbox{CDF-N} survey and review the prospects for deeper {\it Chandra}
exposures. $H_{0}=65$~km~s$^{-1}$~Mpc$^{-1}$, $\Omega_{\rm
M}={1\over3}$, and $\Omega_{\Lambda}={2\over3}$ are adopted
throughout.

%
\section{Observations and Basic Source Properties}\label{data}
%

The CDF-N observations were obtained with ACIS-I (the imaging array of
the Advanced CCD Imaging Spectrometer; Garmire et~al. 2002). The 20
observations that comprise the full dataset were taken over a 27 month
period; the total exposure time is 1.945~Ms. Each observation was
centred close to the Hubble Deep Field-North (HDF-N; Williams
et~al. 1996); the total area of the combined observations is
$\approx460$~arcmin$^2$. The X-ray data processing was similar to that
described in Brandt et~al. (2001a) for the 1~Ms {\it Chandra}
exposure.

Five hundred and three (503) X-ray sources are detected with a {\sc
wavdetect} false-positive probability threshold of 10$^{-7}$ down to
on-axis 0.5--2.0~keV (soft-band) and \hbox{2--8~keV} (hard-band) flux
limits of $\approx1.5\times10^{-17}$~erg~cm$^{-2}$~s$^{-1}$ and
$\approx1.0\times10^{-16}$~erg~cm$^{-2}$~s$^{-1}$,
respectively.\footnote{We searched for X-ray sources in the full,
soft, and hard X-ray bands in addition to four sub bands (0.5--1,
1--2, 2--4, and 4--8~keV).} The adaptively smoothed 0.5--8.0~keV
(full-band) 2~Ms {\it Chandra} image is shown in Figure~1. The
$<$~1$^{\prime\prime}$ positional uncertainty for the majority of the
X-ray sources allows for accurate cross-correlation to
multi-wavelength counterparts.

\begin{figure}
\resizebox{\hsize}{!}
{\includegraphics[]{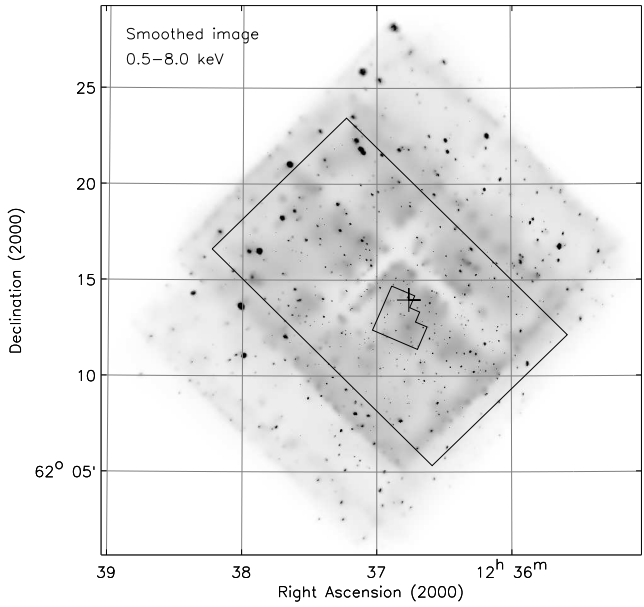}}
\caption{Full-band adaptively smoothed ($2.5\sigma$) image of the CDF-N. 
Much of the apparent diffuse emission is instrumental background, and
the light grooves running through the images correspond to the CCD
gaps. The source sizes change across the field due to the spatial
dependence of the PSF. The small polygon indicates the HDF-N, the
large rectange indicates the GOODS area, and the cross indicates the
average aim point, weighted by exposure time.}
\label{figlabel}
\end{figure}

The range of optical magnitudes for the 503 X-ray detected sources is
huge, spanning $I\approx$~13--26 (note that $\approx$~10\% of the
sources are optically blank). Two hundred and twenty-four (224) of the
sources have redshifts in the literature (e.g.,\ Cohen et~al. 2000;
Barger et~al. 2002), corresponding to $\approx$~45\% of the entire
sample (see Figure~2). Broad-line AGNs (BLAGNs) are detected out to
high redshift (the $z=$~5.186 AGN is the highest-redshift X-ray
selected source known; Barger et~al. 2002). Higher redshift AGNs may
be detected; however, since Lyman-$\alpha$ leaves the $I$-band at
$z{\lower.5ex\hbox{$\;
\buildrel > \over \sim \;$}}6$, many will be optically blank (see
Alexander et~al. 2001 for basic constraints). On average, the $I<23$
non-BLAGN sources follow the redshift track expected for an $L_*$
galaxy (presumably the host galaxy dominates the optical emission);
the statistics are poor for the $I=$~23--24 sources.

Very few $I\ge24$ X-ray sources have redshift constraints. However,
assuming that they follow the same trend as found for the $I<23$
sources, they are likely to lie at $z\approx$~1--3 on average (see
Alexander et~al. 2001 for further discussion). Since these optically
faint X-ray sources could account for $\approx$~50\% of the AGNs
detected in deep {\it Chandra} surveys (e.g.,\ Alexander
et~al. 2002c), determining their redshifts will be crucial for AGN
studies. Until the advent of 30 metre-class telescopes and {\it NGST},
photometric redshifts will probably be the most practical method of
redshift determination for the majority of these sources. The deep
{\it HST} and ground-based observations obtained as part of the Great
Observatories Origins Survey (GOODS) should make this possible for
many of the X-ray sources in the CDF-N and CDF-S surveys.\footnote{See
http://www.stsci.edu/science/goods/ for details on GOODS.}

\begin{figure}
\resizebox{\hsize}{!}
{\includegraphics[]{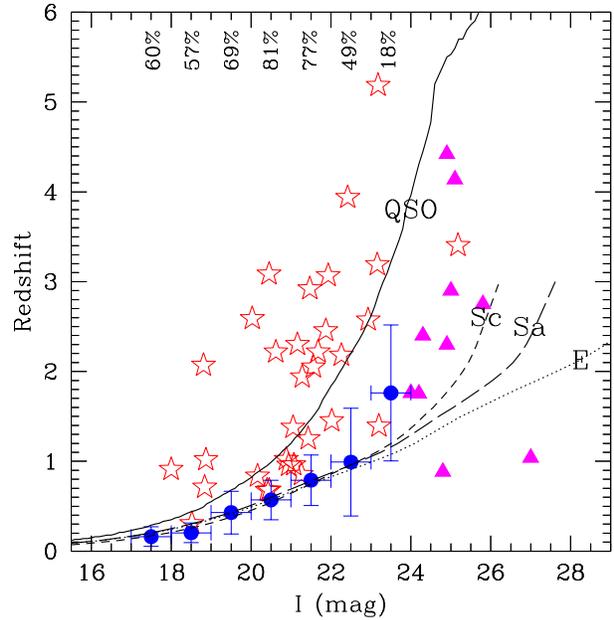}}
\caption{Redshift versus $I$-band magnitude for the X-ray sources with known redshifts. The percentages shown at the top correspond to the fraction of $I=$~17--24 sources with redshifts. The filled circles indicate the mean redshifts for the non-BLAGN sources and the open stars show the redshifts for the BLAGNs. The filled triangles show individual redshifts for the few $I\ge$~24 sources with constraints. The curves show the redshift tracks for $L_*$ galaxies and an $M_I=$~--23 BLAGN (see Alexander et~al. 2001 for more details).}
\label{figlabel}
\end{figure}

%
\section{The Diversity of X-ray Selected Sources}\label{irx}
%

A broad variety of source populations is detected in deep {\it
Chandra} surveys. Here we investigate the diversity of the X-ray
selected sources by focusing on the 21.5~arcmin$^2$ region covered by
the deep 15~$\mu$m ISOCAM observations of Aussel et~al. (1999). In
addition to the ISOCAM observations, this region contains the most
sensitive X-ray and optical observations (i.e.,\ the HDF-N; Williams
et~al. 1996). The $I$-band magnitude versus full-band flux of the 72
X-ray sources detected in this region is shown in Figure~3.

\subsection{AGNs}

Following Alexander et~al. (2002a) we identify AGNs as those sources
with a ``Q'' (i.e.,\ BLAGN) optical spectral classification from Cohen
et~al. (2000), $L_{\rm 0.5-8.0~keV}>3\times10^{42}$ erg~s$^{-1}$, or a
flat effective X-ray spectral slope (i.e.,\ $\Gamma<1.0$, an indicator
of obscured AGN activity).\footnote{All X-ray luminosities are
calculated in the rest frame conservatively assuming $\Gamma=2.0$ and
no intrinsic or Galactic absorption.}  Clearly there can be
AGN-dominated sources not selected by these conserative criteria;
however, the sources classified here as AGNs should be secure. The
AGNs generally fall along the AGN locus defined by Maccacaro
et~al. (1988) from {\it Einstein} observations, even though they are
$\approx$~4 orders of magnitude fainter.

A broad variety of AGNs is found. Very luminous AGNs (i.e.,\ $L_{\rm
0.5-8.0~keV}>10^{44}$ erg~s$^{-1}$) such as QSOs and obscured QSOs are
detected: the former are quite numerous (many of the BLAGNs in
Figure~2 are QSOs), while many of the latter may be optically faint
and difficult to identify (e.g.,\ the $I=$~25.8 obscured QSO in the
HDF-N; Brandt et~al. 2001b). Moderately luminous AGNs such as Seyfert
galaxies are detected out to high redshift. The highest redshift
Seyfert galaxy identified to date is a $z=$~4.424 AGN located close to
the HDF-N (Waddington et~al. 1999; Brandt et~al. 2001b). This source
has $L_{\rm 0.5-8.0~keV}\approx2\times10^{43}$ erg~s$^{-1}$; see
Vignali et~al. (2002 and this volume) for X-ray spectral analysis of
the three $z>4$ AGNs in the CDF-N. However, the most commonly detected
AGNs are moderate-to-low luminosity sources at $z{\lower.5ex\hbox{$\;
\buildrel < \over \sim \;$}}1$. These AGNs show a variety
of characteristics, including sources with emission-line dominated
spectra and sources with absorption-line dominated spectra (e.g.,\
Hornschemeier et~al. 2001; Brandt et~al. 2001b).

X-ray spectral analysis of the X-ray brightest AGNs are presented in
Bauer et~al. (this volume). Amongst other things they report a
diversity in the X-ray spectral properties of the AGNs and show that
the classical view of BLAGNs having unobscured X-ray continua may not
hold at the X-ray fluxes of this survey.

\begin{figure}
\resizebox{\hsize}{!}
{\includegraphics[]{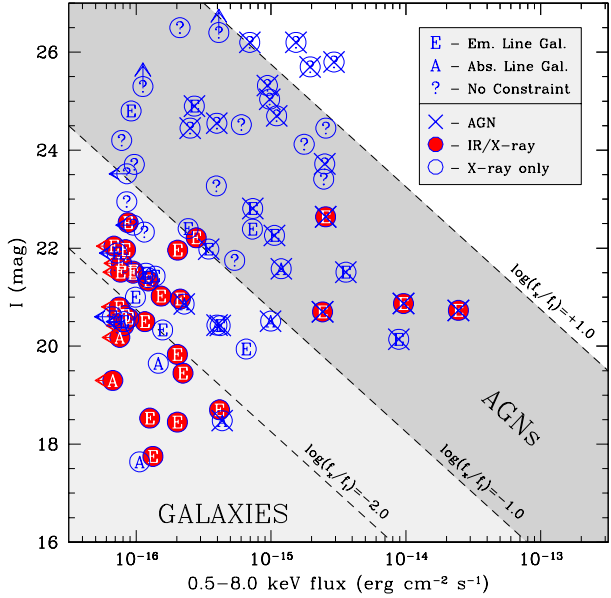}}
\caption{$I$-band magnitude versus full-band flux for the 72 X-ray sources detected in the ISOCAM region. The diagonal lines indicate constant flux ratios. The shaded regions show the approximate flux ratios of the AGN-dominated sources and galaxies (both starbursts and normal galaxies). The classifications of the X-ray sources are indicated as shown in the key.}
\label{figlabel}
\end{figure}

\subsection{Starbursts and normal galaxies}

A large number of apparently normal galaxies are detected at faint
X-ray fluxes (i.e.,\ full-band fluxes
$<5\times10^{-16}$~erg~cm$^{-2}$~s$^{-1}$; Hornschemeier
et~al. 2001). A good indication of the nature of these sources is
found by cross-correlating them with 15~$\mu$m ISOCAM sources. In the
2~Ms dataset we find that 29 of the 40 ($\approx$~73\%) ISOCAM
galaxies in the complete ($f_{\rm 15\mu m}\ge$~100~$\mu$Jy) sample of
Aussel et~al. (in preparation) have X-ray counterparts.\footnote{The
matching was performed to sources detected in any of the seven X-ray
bands; 26 sources are detected in at least one of the three main
bands.} The majority ($\approx$~80\%) of the X-ray detected 15~$\mu$m
sources appear to be emission-line galaxies (ELGs) without apparent
AGNs; by comparison, AGNs and absorption-line galaxies comprise
$\approx$~15\% and $\approx$~5\%, respectively. Of course it is
possible that some of the X-ray detected ELGs contain an AGN not
identified by our conservative criteria; however, their
stacked-average X-ray spectral slope ($\Gamma\approx$~2.0) suggests
that few obscured AGNs are likely to be present. The most likely
scenario is that these sources are predominantly luminous infrared
starburst galaxies (e.g.,\ Alexander et~al. 2002a), a conclusion
entirely consistent with analyses of the faint 15~$\mu$m ISOCAM source
population (e.g.,\ Chary
\& Elbaz 2001).

The range of redshifts and X-ray luminosities for these sources is
broad ($z=$~0.078--1.275 and $L_{\rm 0.5-2.0~keV}=10^{39}$--$10^{42}$
erg~s$^{-1}$). While the X-ray emission from the low-redshift,
low-luminosity sources could be produced by a single ultra-luminous
X-ray source (e.g.,\ Hornschemeier et~al. 2002b), the majority of
these sources have X-ray luminosities between those of M~82 and
NGC~3256, implying moderate-to-luminous star-formation activity
(i.e.,\ $\approx$~10--100 $M_{\sun}$). The two most luminous sources
have X-ray luminosities $\approx$~3 times greater than that of
NGC~3256, the most X-ray luminous starburst galaxy known. The steep
X-ray spectral slopes of these two sources ($\Gamma>1.5$ and
$\Gamma>1.8$) suggests that neither are obscured AGNs.

Most of the X-ray detected ELGs are also detected at radio wavelengths
(Bauer et~al. 2002), providing additional support for the starburst
galaxy hypothesis. Indeed, the X-ray and radio luminosities of these
sources are correlated and agree with those found for local starburst
galaxies, suggesting that the X-ray emission can be used directly as a
star-formation indicator (see also Ranalli, this volume).

%
\section{Prospects for Deeper Chandra Exposures}\label{irx}
%

In examining the prospects for deeper {\it Chandra} exposures we first
review the observations within the extremely sensitive 5.3~arcmin$^2$
\hbox{HDF-N} region (e.g.,\ Williams et~al. 1996). All of the 20 X-ray
sources in the HDF-N have optical counterparts down to $I\approx$~28
and all have spectroscopic or photometric redshifts (see Figure~4);
further sources can be found by matching lower-significance {\it
Chandra} sources to optically bright galaxies (e.g.,\ Figure~5 of
Brandt et~al. 2002).

\begin{figure}
\resizebox{\hsize}{!}
{\includegraphics[]{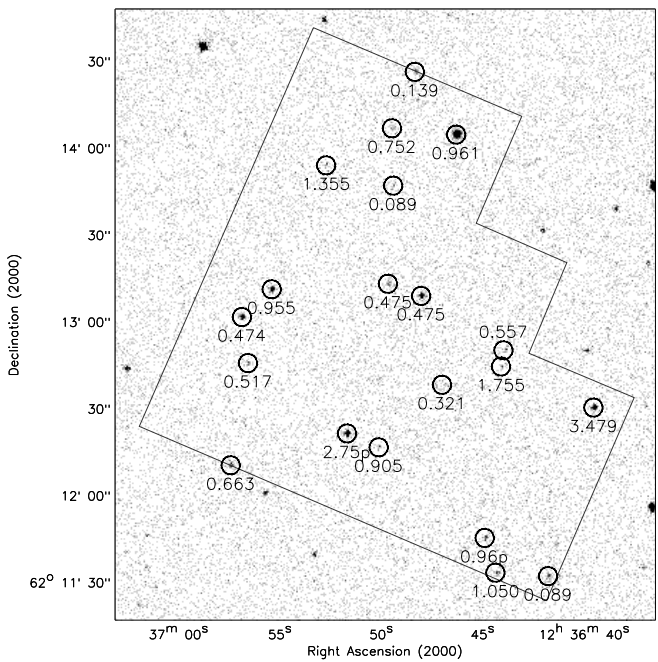}}
\caption{Full-band {\it Chandra} image of the HDF-N and its immediate 
environs. The polygon indicates the HDF-N, and the circles indicate the
X-ray detected sources. Source redshifts are shown above or below each
source (p indicates a photometric redshift). See
http://www.astro.psu.edu/users/niel/hdf/hdf-chandra.html for a movie
of the HDF-N for $\approx$~25--1945~ks {\it Chandra} exposures.}
\label{figlabel}
\end{figure}

The density of X-ray sources in the HDF-N is $\approx$~12,000
deg$^{-2}$ and $\approx$~6,000 deg$^{-2}$ for the soft and hard bands,
respectively. These are the highest source densities reported for a
blank-field X-ray survey, although they are low when compared to the
optical source density of the HDF-N ($\approx$~2~$\times
10^{6}$~deg$^{-2}$; e.g.,\ Williams et~al. 1996). Since the
(conservative) source confusion limit for on-axis {\it Chandra}
observations is $\approx$~80,000~deg$^{-2}$, source confusion is
unlikely to start becoming a problem until significantly fainter X-ray
flux limits are reached (the source confusion limit is calculated
assuming 20~beams per source and a 1.5 arcsec 95\% encircled energy
radius). The average $\approx$~2~Ms {\it Chandra} background is
$\approx$~0.4 counts pixel$^{-1}$ in the 0.5--8.0~keV band, suggesting
that {\it Chandra} should remain close to photon limited for exposures
up to $\approx$~5~Ms, on axis; the situation is even more
favourable in the X-ray bands with lower background rates (e.g.,\ the
0.5--2.0~keV band). Clearly deeper {\it Chandra} exposures can reach
significantly deeper flux limits efficiently and be free of source
confusion.

Several analyses have provided {\it direct} evidence that deeper {\it
Chandra} exposures will detect a significantly larger number of
sources. For instance, X-ray stacking analyses of normal galaxies have
suggested that $\approx$~$L_*$ galaxies at $z<1.5$ should be detected
in significant numbers with 0.5--2.0~keV fluxes of
$\approx$~(3--6)~$\times 10^{-18}$~erg~cm$^{-2}$~s$^{-1}$ (e.g.,\
Hornschemeier et~al. 2002a).\footnote{Stacking analyses of Lyman-break
galaxies and VROs/EROs have also suggested that the average source
should be detectable at these fluxes (e.g.,\ Brandt et~al. 2001c;
Alexander et~al. 2002b).} These results are bolstered by X-ray
fluctuation analyses and model simulations which have implied a rise
in the number counts around 0.5--2.0~keV fluxes of $\approx$~7~$\times
10^{-18}$~erg~cm$^{-2}$~s$^{-1}$ (Miyaji \& Griffiths 2002; Ptak
et~al. 2001). A $\approx$~5~Ms {\it Chandra} survey can achieve on
axis 0.5--2.0~keV fluxes of $\approx$~6~$\times
10^{-18}$~erg~cm$^{-2}$~s$^{-1}$.

A 5--10~Ms {\it Chandra} exposure can reach the flux limits being
discussed for the next generation of X-ray observatories such as {\it
XEUS} (not to be launched for 10--15 years). Although the prime
scientific focus of {\it XEUS} is X-ray spectral analysis, an
ultra-deep {\it Chandra} survey can explore the likely source
populations to be detected in deep {\it XEUS} surveys, and provide
firm constraints on the X-ray source density.

%
\acknowledgements
%
This work has been largely possible due to the financial support of
NASA grants NAS~8-38252 and NAS~8-01128, the NSF CAREER award
AST-9983783, and CXC grant G02-3187A.

\end{document}